\documentclass[twocolumn,aip,apl,reprint, groupedaddress,preprintnumbers,amsmath,amssymb]{revtex4-1}

\usepackage{graphicx}% Include figure files
\usepackage{dcolumn}% Align table columns on decimal point
\usepackage{bm}% bold math
\usepackage{bibtopic}

\usepackage{lipsum}% http://ctan.org/pkg/lipsum
\usepackage{fancyhdr}% http://ctan.org/pkg/fancyhdr
\fancypagestyle{title}{%
  \setlength{\headheight}{22pt}%
  \fancyhf{}% No header/footer
  \fancyfoot[C]{\thepage}% Page number in Centre of footer
  \fancyhead[L]{\small\begin{tabular}{@{}l}
 \scriptsize
 The following article has been submitted to Applied Physics Letters. If it is published, it will be found online at http://apl.aip.org 
 \footnotesize
  \end{tabular}}
}%

\begin{document}

%\preprint{APS}
%\draft
%

\newenvironment{myfont}{\fontfamily{ccr}\selectfont}{\par}

\def\Ef{$E_{\rm F}$}
\def\Eb{$E_{\rm B}$}
\def\Tc{$T_{\rm C}$}
\def\kpara{{\bf k}$_\parallel$}
\def\kperp{{\bf k}$_\perp$}
\def\dirGK{$\overline{\rm \Gamma}\overline{\rm K}$}
\def\dirGM{$\overline{\rm \Gamma}\overline{\rm M}$}
\def\dirGS{$\overline{\rm \Gamma}\overline{\rm S}$}
\def\dirGN{$\overline{\rm \Gamma}\overline{\rm N}$}
\def\dirGNhalf{$0.4\overline{\rm \Gamma}\overline{\rm N}$}
\def\dirGNhalfTH{$0.38\overline{\rm \Gamma}\overline{\rm N}$}
\def\pntG{$\overline{\rm \Gamma}$}
\def\pntM{$\overline{\rm M}$}
\def\pntK{$\overline{\rm K}$}
\def\pntN{$\overline{\rm N}$}
\def\pntNhalf{$0.4\overline{\rm \Gamma}\overline{\rm N}$}
\def\pntNfourty{$0.4\overline{\rm \Gamma}\overline{\rm N}$}
\def\pntNfourtyTH{$0.38\overline{\rm \Gamma}\overline{\rm N}$}

\def\invA{\AA$^{-1}$}
\def\DCgamma{${\rm DC}_{\overline{\Gamma}}$}
\def\DCNhalf{$  {\rm DC}_{0.4\overline{\rm \Gamma}\overline{\rm N}} $}

%\def\DCgamma{$\Gamma$}
%\def\DCNhalf{$N/2$}

%\title{Suppression of electron scattering resonances in graphene by\\ 
%nanopatterning with quantum dots}

%\title{Suppression of electron scattering resonances in graphene by\\
%superlattice of quantum dots}

%\newcommand\PlaceText[3]{%
%\begin{tikzpicture}[remember picture,overlay]
%\node[outer sep=0pt,inner sep=0pt,anchor=south west] 
%  at ([xshift=#1,yshift=-#2]current page.north west) {#3};
%\end{tikzpicture}%
%}

\title{Suppression of electron scattering resonances in graphene by quantum dots}

\author{M. Krivenkov, D. Marchenko, J. S\'anchez-Barriga, O. Rader and A. Varykhalov}

\affiliation{Helmholtz-Zentrum Berlin f\"ur Materialien und Energie, 
Elektronenspeicherring BESSY II, Albert-Einstein-Str. 15, D-12489 Berlin, Germany}

\begin{abstract}

Transmission of low-energetic electrons through two-dimensional materials 
leads to unique scattering resonances. These resonances contribute to photoemission 
from occupied bands where they appear as strongly dispersive features of suppressed 
photoelectron intensity. Using angle-resolved photoemission we have systematically 
studied scattering resonances in epitaxial graphene grown on the chemically differing 
substrates Ir(111), Bi/Ir, Ni(111) as well as in graphene/Ir(111) nanopatterned with 
a superlattice of uniform Ir quantum dots. While the strength of the chemical interaction 
with the substrate has almost no effect on the dispersion of the scattering resonances, 
their energy can be controlled by the magnitude of charge transfer from/to graphene. 
At the same time, a superlattice of small quantum dots deposited on graphene eliminates 
the resonances completely. We ascribe this effect to a nanodot-induced buckling of graphene 
and its local rehybridization from {\it sp}$^{2}$ to {\it sp}$^{3}$ towards a 
three-dimensional structure. Our results suggest nanopatterning as a prospective tool 
for tuning optoelectronic properties of two-dimensional materials with graphene-like structure.
\end{abstract}
 
%\pacs{}  

\maketitle
\thispagestyle{title}

%\PlaceText{12mm}{-5mm}{
%\begin{myfont}
%\small
%The following article has been submitted to Applied Physics Letters. If it is published, it will be found online at http://apl.aip.org
%\end{myfont}
%}

Scattering resonances in the unoccupied band structure of two-dimensional (2D) materials 
is a unique phenomenon predicted theoretically several years ago.\cite{Nazarov-ScR} 
In terms of band structure they resemble narrow band gaps 
dispersing in a wide range of electron energies, starting from 
nearly the vacuum level and propagating for several tens of electronvolts 
above. In the region of such resonances, 2D materials become 
nontransparent for incoming electrons as electron reflectivity 
can reach 100\%. 
From a quantum-mechanical point of view these resonances are born by 
the coupling of in-plane and out-of plane motions of electrons in the
2D material,\cite{Nazarov-ScR} which is permitted by the
in-plane periodic potential of the crystal lattice. Scattering resonances have 
an essential impact on optoelectronic applications of 2D materials and 
especially of graphene.\cite{Nazarov-ScR}

These resonances can be naturally described in terms of a time-reversed low-energy electron diffraction (LEED) process. Correspondingly, angle-resolved very-low-energy electron diffraction (VLEED) can be used \cite{Strocov-VLEED} to probe them but with limited success. VLEED does not
resolve band dispersions precisely but only boundaries of the surface-projected bands.\cite{Strocov-VLEED} Certain 
improvements in the visualization of their dispersions were achieved for VLEED-based low-energy electron microscopy only recently.\cite{Jobst-LEEM} An alternative method is electron holography allowing to directly observe angle-resolved scattering patterns of electrons emitted from a point source and trespassing graphene.\cite{Wicki-Fink-Holo} 
This method, however, works only for freestanding suspended graphene.
Yet another and powerful technique is angle-resolved photoelectron 
spectroscopy (ARPES). Here graphene is prepared on a substrate and
irradiated with photons, and the angular scattering of secondary 
electrons emitted from the substrate and passing through the 
graphene overlayer is acquired (principally, a similar variation of this method is angle-resolved secondary-electron emission \cite{Maeda-ARSEE, Pisarra-ARSEE}). In spite of the fact that ARPES can only access the occupied band structure, the scattering resonances, hosted above Fermi level, can be observed using ARPES. This is due to the coupling between occupied and unoccupied bands in the final states of photoemission.\cite{Mahatha-PRB} Furthermore, the kinetic energy of the graphene resonances is independent of the energy of the exciting photons.\cite{Mahatha-PRB}

In the present Letter, we address the question of
how scattering resonances of enhanced electron reflectivity 
from graphene can be tuned, modified or suppressed either by
chemical interaction with the substrate or by enhancement of 
the lateral modulation of the graphene crystal potential by a superlattice 
of quantum dots. In particular, using ARPES, we study epitaxial 
graphene weakly bonded to Bi/Ir(111), moderately bonded 
to Ir(111) and strongly bonded to Ni(111), as well as graphene on Ir(111) 
nanopatterned with an array of uniform Ir dots. 
We demonstrate that the dispersions of the scattering resonances are
robust toward the grade of interaction between graphene and the 
substrate, with the exception that their kinetic energies consistently follow 
the magnitude of charge transfer from/to graphene. Moreover, we reveal 
these resonances to be extremely sensitive to nanodots residing 
on graphene. The resonances disappear upon deposition of even very small Ir 
dots of $\sim$ 0.1 monolayer (ML) nominal thickness. This effect is explained by nanodot-induced rehybridization of the graphene structure.

Figure 1 shows how the scattering resonances of graphene \cite{Nazarov-ScR} 
occur in the ARPES spectra. The exemplary data was measured from weakly-bonded quasi-freestanding graphene on Bi (prepared as graphene/Ir(111) and subsequently intercalated with 1 ML Bi) using a photon energy of h$\nu$=80 eV. Figure 1(a) displays energy-momentum dispersions measured along the \dirGK\ direction of the graphene surface Brillouin zone (SBZ) and over a wide range of kinetic energies, ranging from 13 eV (secondary electrons) to 78 eV (just above Fermi level). Since visualizing the resonances in the raw data is difficult due to the high intensity of secondary electrons, here we show the derivative of the intensity $\frac{dI}{dE}$. Various electronic states of graphene and the buried Ir substrate, mostly seen at high kinetic energies, are denoted with the labels {\it Gr} and {\it Ir}, respectively.

At lower kinetic energies ($<$50 eV), we observe broad bright peaks 
with crossing parabolic dispersions (denoted as {\it FrE}) 
which are free-electron bands of the vacuum continuum
backfolded with the Brillouin zone periodicity.\cite{Pisarra-ARSEE}
In addition, there are very sharp narrow features (denoted as {\it ScR}) 
with parabolic dispersions that cross at the \pntG\
point of the SBZ and at a kinetic energy of $\sim$32 eV. 
These features are the graphene scattering resonances under discussion. To provide an impression on how they appear in the raw data, we refer to Fig. 1(g). Here the resonances are seen as pronounced intensity dips due to the nontransparency of graphene to the photoelectrons.  Along \dirGM\ 
[data not shown] the resonances exhibit a split structure in full agreement with the theory.\cite{Nazarov-ScR} Both free electron bands {\it FrE} and scattering 
resonances {\it ScR} are hosted above Fermi level, but 
become observable in the ARPES signal due to couping with the final state of photoemission. 
In consequence, their kinetic energies {\it persist} and do not depend on the excitation energy of the photons.

Figures 1(b)-1(e) [left panels] show a sequence of constant-kinetic-energy
surfaces obtained from full photoemission mapping of the scattering resonances {\it ScR} (made for the first time using ARPES), revealing their localization within the Brillouin zone. Right panels in Figs. 1(b)-1(e) display the dispersions of both {\it ScR} (black lines) and free-electron bands {\it FrE} (green lines) as extracted from the data shown on the left panels. The shape of the {\it ScR} patterns varies from hexagonal towards a 'star of David', in good agreement with results of electron holography experiments on freestanding graphene.\cite{Wicki-Fink-Holo} The overall band structure of the scattering resonances can be clearly understood from the ARPES maps. It is composed of six arcs which are the parts of six backfolded parabolas centered at the \pntG\ points of the neighbouring Brillouin zones. Exactly the same construction holds for the free-electron bands {\it FrE}. Thus, the dispersion of {\it ScR} in {\bf k}-space closely follows the one of {\it FrE}, in full agreement with the expected behavior of non-tight-binding bands predicted theoretically.\cite{Kogan-PSS}

Figures 1(f) and 1(g) emphasize the remarkable impact that the scattering resonances have on the optoelectronic properties of graphene. Here we show the valence band of graphene/Bi/Ir(111) containing the dispersions of $\sigma_{2,3}$ and $\pi$ states measured near the \pntG\ of the SBZ at h$\nu$=62 eV and h$\nu$=35 eV, respectively. While at 62 eV both $\sigma$ and $\pi$ bands look undistorted, at 35 eV they overlap with the scattering resonances {\it ScR} (the kinetic energy of which does not depend on photon energy), and become severely modified by acquiring artificial band gaps. Markedly, the scattering resonances suppress not only the secondary electrons from the Bi/Ir substrate trespassing graphene, but also the signal from $\sigma$ and $\pi$ bands which are emitted from graphene itself.

\begin{figure}[t]
\centering
\includegraphics[width=0.48\textwidth]{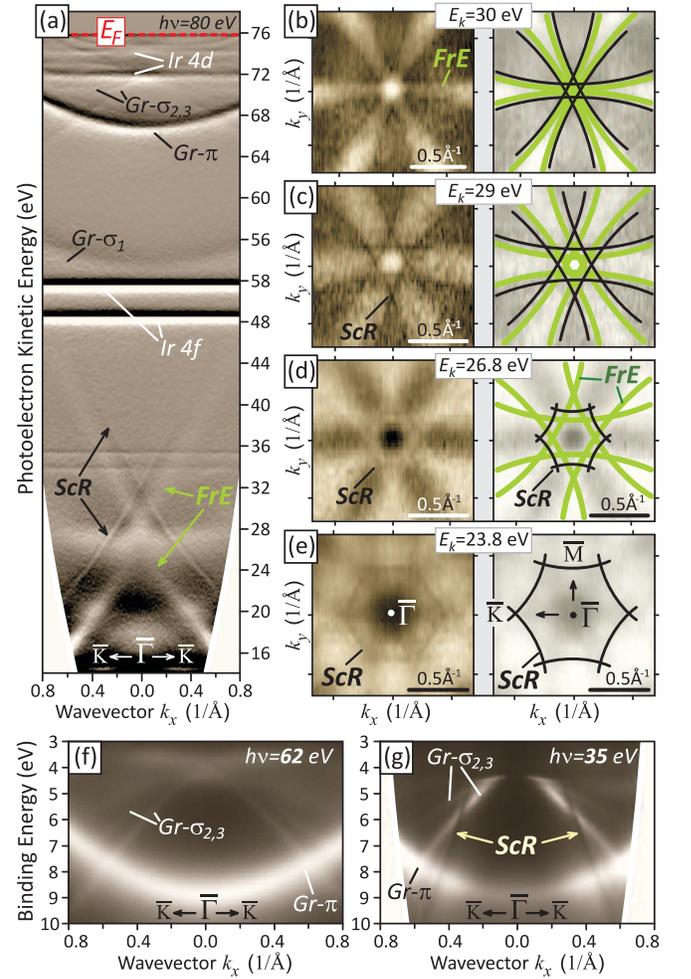}
\caption{Scattering resonances {\it ScR} and free-electron final state bands {\it FrE} in graphene/Bi revealed by ARPES. (a) ARPES spectrum along \dirGK\ shown as the first derivative of the intensity ($\frac{dI}{dE}$), where parabolic dispersions of {\it ScR} crossing at a kinetic energy of $\sim$ 32 eV are seen. (b)-(e) Sequence of constant-kinetic-energy surfaces of scattering resonances {\it ScR}. The arcs of {\it ScR} [black lines]
closely follow the backfolded free-electron parabolas {\it FrE} [green lines].
(f),(g) Raw photoemission data from $\sigma_{2,3}$ and $\pi$ bands of graphene measured along \dirGK\ at (f) 62 eV and (g) 35 eV photon energy.}
\end{figure}

\begin{figure}[t]
\centering
\includegraphics[width=0.48\textwidth]{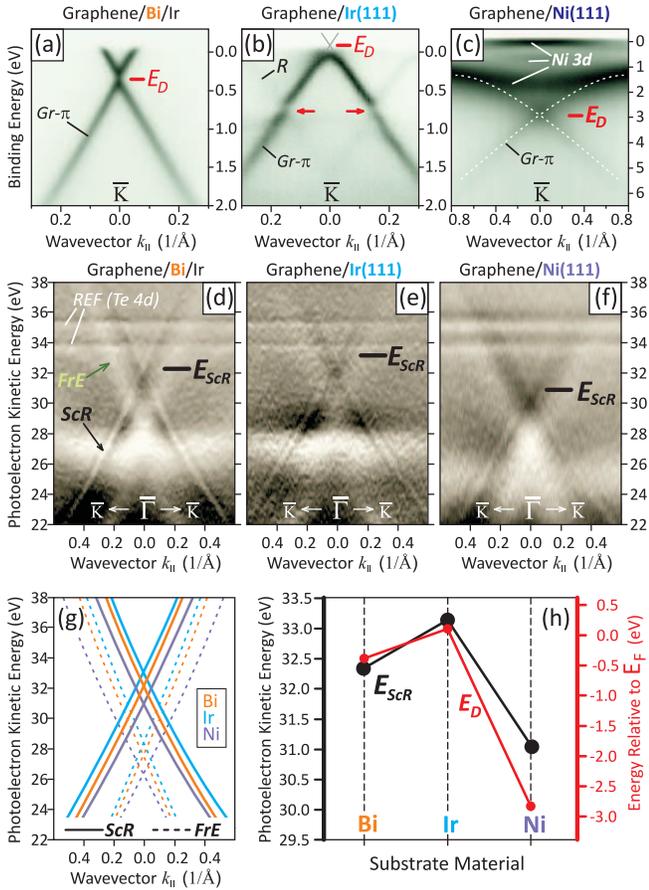}
\caption{
Effect of various substrates on the graphene scattering resonances {\it ScR}.(a)-(c) The Dirac cones in graphene on (a) Bi/Ir, (b) Ir and (c) Ni reveal 
different interaction strengths between graphene and substrate.
(d)-(f) For all three substrates, ARPES reveals the presence of scattering resonances {\it ScR} and free-electron bands {\it FrE} which (g) slightly differ in their kinetic energy. (h) The magnitude of the energy shift correlates with the charge-transfer between graphene and its substrate, as extracted from (a)-(c).}
\end{figure}

\begin{figure}[t]
\centering
\includegraphics[width=0.48\textwidth]{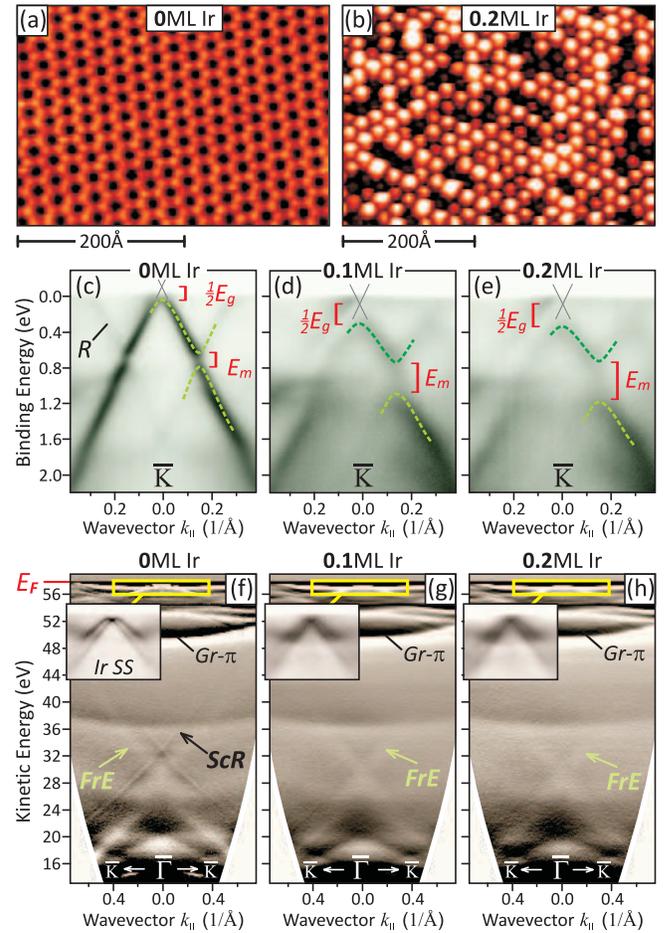}
\caption{
Impact of Ir quantum dots deposited on graphene/Ir(111) on the scattering
resonances {\it ScR}. (a)-(b) STM characterization of (a) bare moir\'e-patterned graphene/Ir(111) and (b) deposited with 0.2 ML of Ir self-assembled into quantum dots. (c)-(e) Ir dots enlarge with increasing concentration of Ir and
enhance the lateral modulation of moir\'e superpotential, which increases the size of the minigaps {\it E$_m$}. (f)-(h) Behavior of the scattering resonances {\it ScR} with the enhancement of the superpotential by the dots. The resonances are suppressed by even very small Ir dots while the free-electron bands {\it FrE} remain unaffected.}
\end{figure}

To investigate whether the scattering resonances can be manipulated, we have tested their behavior in graphene grown on various substrates with contrasting chemical properties that differently affect graphene. In particular, we have
studied graphene on Bi [prepared by intercalation of 1 ML Bi under graphene on Ir(111)], pristine graphene/Ir(111) and graphene/Ni(111). The interaction strength between graphene and its substrate can be judged from the behavior of the Dirac cone at the \pntK\ point of the SBZ [Figs. 2(a)-(c)]. 

The Dirac cone in graphene/Bi [Fig. 2(a)] appears almost intact with a very minor band gap and rather small charge doping ($\sim$390 meV). Also, no electronic hybridization between graphene and the Bi layer forms,\cite{Shikin-NJP} which means that graphene is quasifreestanding and nearly undisturbed by the Bi.
In graphene/Ir(111) [Fig. 2(b)], the Dirac cone exhibits a band gap at the Dirac point \cite{Sanchez-Barriga-Clusters,Carbone-Clusters} 
close to the Fermi level, as well as renowned band replicas and minigaps (red arrows), which are born from {\it umklapp} scattering due to the moir\'e superpotential.\cite{Pletikosic-minigaps} Considering the weak electronic hybridization between graphene and Ir {\it 4d} bands,\cite{Marchenko-PRB-Ir} the graphene-substrate interaction is judged in this case as of moderate grade. In contrast, as shown in Fig. 2(c), the Dirac cone in graphene/Ni(111)
\cite{Varykhalov-PRX} is severely modified by the covalent interaction with the substrate \cite{Shikin-covalent} and by the strong hybridization with the Ni {\it 3d} bands.\cite{Varykhalov-PRX} This is an example of strongly bonded graphene.

Surprisingly, the graphene scattering resonances {\it ScR} observed for all three systems are {\it robust} against the chemical interaction between graphene and its substrate. Their appearance in graphene/Bi, graphene/Ir(111) and graphene/Ni(111) is evidenced by the ARPES data shown in Figs. 2(d), 2(e) and 2(f), respectively. In Fig. 2(g), we represent the dispersions of the scattering resonances and free-electron bands {\it FrE}) as extracted precisely from the ARPES data shown in Figs. 2(d)-2(f). Apart from the clear energy shifts, also seen directly in Figs. 2(d)-2(f), the energy-momentum dispersion of the graphene resonances is {\it identical} in the three systems. The magnitude of these energy shifts was evaluated carefully by using the {\it 4d} core levels of very 
small amounts of Te adatoms as additional reference for the energy calibration (denoted as {\it REF}). In Fig. 2(h), we show the energy positions of {\it ScR} relative to
their crossing energy {\it E$_{ScR}$} for Bi, Ir and Ni substrates. The values correlate very well with the amount of charge transfer from/to graphene, as determined from Figs. 2(a)-2(c) by the energy position of the original Dirac points ({\it E$_D$}), which is also plotted in Fig. 2(h). The somewhat larger discrepancy for the Ni substrate can be explained by 
an enhanced contribution
of the covalent graphene-Ni bonding to the shift and the concomitant scaling
of the band structure.\cite{Shikin-covalent} 
%In addition, 
One should also consider that the Brillioun zone of graphene/Ni, as compared to the one of 
graphene/Ir, is a bit smaller due to the slightly enlarged lattice constant.\cite{Tamtoeg-JPC-2015} 
This has an impact on the backfolding of scattering resonances and 
additionally reduces E$_{ScR}$ by up to $\sim$ 1 eV.

According to Ref.~\onlinecite{Nazarov-ScR}, the impact of the scattering resonances on the band structure can be enhanced in corrugated or strained 
graphene because the strength of the in-plane potential modulation influences the quantum mechanical coupling of in-plane and out-of-plane motions of electrons, giving rise to the {\it ScR} resonances. In graphene/Ir(111) there is a renowned moir\'e pattern with a lateral periodicity of $\sim$25\AA\ which occurs due to $\sim$10\%\ lattice mismatch between graphene and Ir.\cite{Pletikosic-minigaps, Marchenko-PRB-Ir} Its STM image is shown in Fig. 3(a). This moir\'e pattern supports self-assembly of room-temperature-stable 
Ir quantum dots. The dots arrange in moir\'e cells, and their sizes are proportional to the concentration of deposited Ir. They locally pin graphene to Ir(111) and strongly enhance its structural corrugation \cite{Feibelman-pinning} which, in turn, increases the superpotential modulation. 
%The nanoparticles are perfectly uniform and arrange in moir\'e cells, 
%while their sizes depend on concentration of deposited material 
%and hence define magnitude of potential modulation.

Therefore, we have tested scattering resonances {\it ScR} in graphene/Ir(111) with a moir\'e superpotential enhanced by quantum dots. Figure 3(b) shows a STM image of graphene/Ir deposited with Ir dots formed from 0.2 ML Ir.
The effect of enhanced structural modulation can be seen in the electronic structure of the graphene Dirac cone.\cite{Sanchez-Barriga-Clusters,Carbone-Clusters}
Figures 3(c), 3(d) and 3(e) show the Dirac cone in bare graphene/Ir
(no dots), in graphene/Ir deposited with 0.1 ML Ir (small dots),
and 0.2 ML Ir (larger dots), respectively. The small dimensions of the deposited Ir dots were also cross-checked by the persistence of the Ir(111) 
surface state \cite{Varykhalov-NJP} under graphene [see insets of Figs. 3(f)-3(h)]. The superpotential modulation induced in graphene by the Ir dots is
seen through enlarged minigaps ({\it E$_m$}) at the crossings with 
the Dirac cone replicas. We obtain {\it E$_m$}$\sim$180 meV for the bare sample, $\sim$320 meV for 0.1 ML Ir and $\sim$420 meV for 0.2 ML Ir. Furthermore, the band gap at Dirac point {\it E$_g$} becomes wider due to the local breaking of the sublattice symmetry in graphene induced by the dots.\cite{Sanchez-Barriga-Clusters}

Remarkably, the graphene scattering resonances {\it ScR} disappear
almost completely upon deposition of even small quantum dots. This is seen in Figs. 3(f), 3(g) and 3(h) which show wide energy range dispersions 
measured along \dirGK\ for bare graphene/Ir(111) and for graphene/Ir decorated with 0.1 ML Ir and 0.2 ML Ir, respectively. To explain this effect, we again refer to Ref.~\onlinecite{Nazarov-ScR}, which discovers that scattering resonances are unique for 2D materials and emerge from the coupling between in-plane and out-of-plane motions of electrons. It is known that an enhancement of the graphene corrugation by Ir quantum dots causes a local rehybridization of graphene from {\it sp}$^{2}$ to {\it sp}$^{3}$ at the sites where the dots reside.\cite{Feibelman-pinning,Knudsen-X-ray} This, in turn, switches graphene from 2D towards a 3D-like structure, lifting the quantum-mechanical coupling between in-plane and out-of-plane electron motions and thus, suppressing the formation of the scattering resonances {\it ScR}. At the same time, the fundamental band structure of graphene, including the Dirac cone, remains persistent [Figs. 3(c)-3(e)].

In summary, we have studied scattering electron resonances occurring in the unoccupied band 
structure of graphene using ARPES. By testing epitaxial graphene on various metallic substrates (Bi, Ir, Ni) 
we have shown that the energy of these resonances can be controlled by the magnitude of charge 
transfer from/to graphene but their dispersions are robust against the strength of the graphene-substrate 
interaction. We have also shown that the scattering resonances can be suppressed completely 
upon decoration of moir\'e patterned graphene/Ir(111) with small Ir quantum dots. This was ascribed 
to a rehybridization of graphene towards {\it sp}$^{3}$ induced by the Ir dots, which lifts the 
quantum mechanical conditions for the formation of the scattering resonances. 
Considering that in general case, graphene locally decorated with small quantum dots 
remains transparent and preserves its fundamental band structure, the 
nanopatterning evolves into a promising method for tuning optoelectronic properties of graphene.

%{\it Acknowledgements.} 
This work was supported by Impuls- und Vernetzungsfonds 
der Helmholtz-Gemeinschaft (Grant No. HRJRG-408).
Authors acknowledge fruitful discussions with E. Kogan 
and A. Fedorov.

\end{document}